\documentclass[english,11pt,aps,showpacs]{revtex4}
\usepackage[T1]{fontenc}
\usepackage[utf8]{inputenc}

\usepackage{color}
\usepackage{babel}
\usepackage{bbold}
\usepackage{array}
\usepackage{longtable}
\usepackage{float}
\usepackage{booktabs}
\usepackage{amsmath}
\usepackage{graphicx}
\usepackage{amssymb}
\usepackage{esint}
\usepackage{feynmf}

\newcommand{\vk}{\textbf{k}}
\newcommand{\vq}{\textbf{q}}
\newcommand{\dd}{{\rm d}}

\usepackage[unicode=true, 
 bookmarks=true,bookmarksnumbered=false,bookmarksopen=false,
 breaklinks=false,pdfborder={0 0 0},backref=false,colorlinks=false]
 {hyperref}
\begin{document}

\title{On the symmetry properties of the PT kernels and recurrence relations}

\author{Paulo Reimberg}
\affiliation{Sorbonne Universit\'es, UPMC Univ Paris 6 et CNRS, UMR
  7095, Institut d'Astrophysique de Paris, 98 bis bd Arago, 75014
  Paris, France \\ \footnotesize{reimberg@iap.fr}}

\begin{abstract}
Perturbation Theory to Large Scale Structure Cosmology proposes
corrections to the linearly evolved density contrast and velocity in
terms of a series development in which all terms are 
integrals of powers of the linear density contrast multiplied by
kernels. We discuss the symmetry properties of these kernels and show
that their full symmetrized versions can be decomposed in different
classes of subkernels. We will construct classes of  subkernels with
improved symmetry properties,  and provide recurrence relations to generate them.
\end{abstract}

\maketitle

Paving the theoretical way that connects primordial cosmological
perturbations produced in inflationary models to the large scale
structure of matter distribution in Universe as observed today is a
major program in theoretical physics. Large scale structure
observables encode information about the underlying cosmological model, but
the connection is hidden by the complex dynamics of gravitational
instability growth, interaction among species and astrophysical
processes.

Simplifications have come to scene so one can start unveiling
structural properties of the gravitational dynamic, and application of
Perturbation Theory to large scale
structure cosmology is one of the best succeeded frameworks allowing
to go beyond the linear theory of gravitational evolution (see
\cite{peebles} or
\cite{review, bernardeau_les_houches} for reviews). The goal in
Perturbation Theory is to evolve primordial fluctuations of density
and velocity for cold dark matter particles by perturbativelly solving
fluid equations that would capture the general properties of a
collisionless gas in a Friedmann-Lamaitre-Robertson-Walker
spacetime. The developments in Perturbation Theory form a theoretical
body on itself but also form the basis for many further frameworks
such as Effective Field Theory approaches to LSS
\cite{baumann2012cosmological, carrasco2012effective}.

Assuming that structure formation will be driven by the action of
gravity on the pressureless fluid of cold dark matter
particles, the basic system of equations to be considered is composed
by continuity, Euler and Poisson equations. Once translated in Fourier
space, these equations become a system of non-homogeneous coupled
non-linear differential equations, for which analytical solutions are
not known. A perturbative solution in a form of series expansion for
density contrast and velocity divergence is then aimed. The general
structure of the terms in the series will be \footnote{We use here the
  results for Einstein-de Sitter spacetimes where the kernels are
  independent of time. This can be generalized and combinatoric
  properties of the kernels are still preserved \cite{Fasiello:2016qpn}}

\begin{displaymath}
\delta^{(n)}_a(\vk) = \int \prod_{i=1}^d \dd \vk_i \, \delta_D \left(\sum_{i=1}^n \vk_i - \vk \right) F_n(\vk_1, \ldots, \vk_n) \prod_{i=1}^n \delta^{(1)}(\vk_j) \, ,
\end{displaymath}
i.e., the $n$-th term in the series expansion is given by a
momentum-conserving integral of $n$ powers of the linear density times
a kernel. The kernel for density and velocity evolutions will be
distinct homogeneous function of the momenta, and symmetric under
permutation of its arguments. This symmetry, however, is obtained
after symmetrization of subkernels obtained in the perturbative
solution for dynamical equations. We will discuss in detail  here the symmetry
properties of these subkernels, and show explicitly how different symmetrization schemes lead to equivalent kernels.

We shall first quickly review the system of equation for which a
perturbative solution is searched, and point out the main two schemes
of perturbative development. We shall then discuss the symmetrization
processes more adapted to each scheme and show, explicitly looking at
low orders, their equivalence. We will argue that the subkernels
generated in the so called \emph{Scoccimarro's method} have better
symmetry properties.  We will present recurrence relations that 
generate the symmetrical kernels at all orders and at a lower computational cost. These kernels will have convenient property of being decomposable into a sum of $\lfloor n/2 \rfloor$ terms symmetric under permutation of the $m$ firsts, and $n-m$ last arguments, $1 \leq m \leq \lfloor n/2 \rfloor$.

\section{Perturbative schemes}

Continuity, Euler and Poisson equations for cold dark matter (CDM) fluid in a FLRW spacetime lead to a system of differential equations describing the fluid's density contrast and velocity. If we take, for instance, the continuity equation, we will have:
\begin{equation}
\label{continuity}
\dot{\delta} + \frac{1}{a} \nabla \cdot \bold{u} = -\frac{1}{a} \nabla \cdot (\delta \bold{u})
\end{equation}
where $a$ is the scale factor on the FLRW metric, $\theta := \nabla
\cdot \bold{u}/(aH)$, and $H= \dot{a}/a$ is the Hubble parameter. On
the right-hand side of Eq. \eqref{continuity} we see the coupling term
between the evolution of density and velocity, namely the divergence
of the product of density and velocity. If we look at the Euler
equation, the same structure appears, but the coupling term consists
of a product of velocities. We do not aim to give full derivation of
the basic equations here, what can be found in \cite{review, RPT, bernardeau_les_houches}, but claim that their structure is:
\begin{equation}
\label{non-hom}
\frac{\partial}{\partial \eta} \Psi_a (\vk, \eta) + \Omega_a^b(\eta) \Psi_b(\vk, \eta) = \int \frac{\dd^3 \vk_1}{(2 \pi)^{3/2}} \int \frac{\dd^3 \vk_2}{(2 \pi)^{3/2}} \gamma_a^{bc}(\vk_1, \vk_2) \Psi_b(\vk_1, \eta) \Psi_c(\vk_2, \eta)
\end{equation}
where latin indexes range from 1 to 2 and are summed if repeated. Here $\eta = \log D_+$, and $D_+$ is the growing mode on linear evolution of gravitational instability. The dublet $\Psi(\bold{x}, \eta) := (\delta(\bold{x}, \eta), -\frac{\theta(\bold{x}, \eta)}{f_+} )$, with $f_+ := \dd \log D_+ / \dd \log a$, encodes the dynamical quantities in the system. The coupling matrix $\Omega(\eta)$ is derived from the linearized equations of motion, and in EdS is:

\begin{equation}
\label{Omega}
\Omega = \left( \begin{array}{cc} 0 & -1 \\ -3/2 & 1/2 \end{array} \right) \, .
\end{equation}

The quantities $\gamma$ are the \emph{vertex coupling}, defined as: 
\begin{equation}
\label{gammas}
\gamma_1^{21}(\vk_1, \vk_2) = \delta_D(\vk - \vk_1 - \vk_2)  \alpha(\vk_1, \vk_2)   \qquad \gamma_2^{22}(\vk_1, \vk_2) = \delta_D(\vk - \vk_1 - \vk_2)  \beta(\vk_1, \vk_2)  
\end{equation}
with
\begin{equation}
\label{alpha_beta}
\alpha(\vk_1, \vk_2) = \frac{(\vk_1 + \vk_2) \cdot \vk_1}{k_1^2}
\qquad
\beta(\vk_1, \vk_2) = \frac{ |\vk_1+\vk_2|^2 \vk_1 \cdot \vk_2}{2 k_1^2 k_2^2} \, .
\end{equation}
We observe that $\vk = \vk_1 + \vk_2$ on Eq. \eqref{non-hom} as
enforced by the Dirac deltas on Eqs. \eqref{gammas}.
The structure of $\gamma$ reproduces the asymmetric couplings in continuity and Euler equations: density couples to velocity on continuity equation, but Euler's equation only couples velocity to velocity. 

There are two standard ways of developing a perturbative solution for Eq. \eqref{non-hom}, referred here as `Goroff's' and `Scoccimarro's' methods, and briefly reviewed in the sequence.

%%%%%%%%%%%%%%%%%%%%%
\subsection{Goroff's method}
%%%%%%%%%%%%%%%%%%%%%
The perturbative solution of Eq. \eqref{non-hom} was developed in a series of papers \cite{peebles, fry, goroff, jain}. We will call the method to be presented here `Goroff's method' because of \cite{goroff}. 

Considering an EdS universe, assume the expansions:

\begin{equation}
\label{delta_orders}
\delta(\bold{x}, t) = \sum_{n=1}^{\infty} \delta^{(n)}(\bold{x}, \eta) = \sum_n a(t)^n \delta^{(n)}(\bold{x})
\end{equation}

\begin{equation}
\label{theta_orders}
\theta(\bold{x}, t) = - \dot{a}(t) \sum_{n=1}^{\infty} a(t)^n \theta^{(n)} (\bold{x}) \, .
\end{equation}
Substituting Eqs. \eqref{delta_orders} and \eqref{theta_orders} in Eq. \eqref{non-hom} we obtain a linear system that determine each $\delta^{(n)}$ and $\theta^{(n)}$. The solutions will clearly depend on the interplay of the vertexes couplings for different momenta at all orders less than $n$ and can be written as \cite{review}:

\begin{equation}
\label{delta_n}
\delta^{(n)}(\vk) = \int \frac{\dd^3 \vk_1}{(2 \pi)^{3/2}} \ldots \int \frac{\dd^3 \vk_n}{(2 \pi)^{3/2}} \delta_D(\vk_1 + \ldots + \vk_n - \vk) F_n(\vk_1, \ldots, \vk_n) \delta^{(1)}(\vk_1) \ldots \delta^{(1)}(\vk_n)
\end{equation}

\begin{equation}
\label{theta_n}
\theta^{(n)}(\vk) = \int \frac{\dd^3 \vk_1}{(2 \pi)^{3/2}} \ldots \int \frac{\dd^3 \vk_n}{(2 \pi)^{3/2}} \delta_D(\vk_1 + \ldots + \vk_n - \vk) G_n(\vk_1, \ldots, \vk_n) \delta^{(1)}(\vk_1) \ldots \delta^{(1)}(\vk_n)
\end{equation}
where $\delta^{(1)}$ is the linear density, and $F_n$, $G_n$ are the \emph{PT kernels}. $F_1=G_1=1$ and, for $n \geq 2$, the kernels are given by the recurrence relations, obtained collecting terms of the same order after substituting Eqs. \eqref{delta_orders} and \eqref{theta_orders} in \eqref{non-hom}:

\begin{eqnarray}
\label{F_recurrence}
F_n(\vk_1, \ldots, \vk_n) & = & \sum_{m=1}^{n-1} \frac{G_m(\vk_1, \ldots, \vk_m)}{(2n+3)(n-1)} \Big[(2n+1) \alpha(\bar{\bold{k}}_1, \bar{\bold{k}}_2) F_{n-m}(\vk_{m+1}, \ldots, \vk_n) \nonumber\\ & & + 2 \beta(\bar{\bold{k}}_1, \bar{\bold{k}}_2) G_{n-m}(\vk_{m+1}, \ldots, \vk_n) \Big] \, ,
\end{eqnarray}

\begin{eqnarray}
\label{G_recurrence}
G_n(\vk_1, \ldots, \vk_n) & = & \sum_{m=1}^{n-1} \frac{G_m(\vk_1, \ldots, \vk_m)}{(2n+3)(n-1)} \Big[3 \alpha(\bar{\bold{k}}_1, \bar{\bold{k}}_2) F_{n-m}(\vk_{m+1}, \ldots, \vk_n) \nonumber\\ & & + 2 n \beta(\bar{\bold{k}}_1, \bar{\bold{k}}_2) G_{n-m}(\vk_{m+1}, \ldots, \vk_n) \Big] \, 
\end{eqnarray}
where $\bar{\bold{k}}_1 := \vk_1 + \ldots + \vk_m$ and $\bar{\bold{k}}_2 := \vk_{m+1} + \ldots + \vk_n$.

%%%%%%%%%%%%%%%%%%%%%%%%%
\subsection{Scoccimarro's method}
%%%%%%%%%%%%%%%%%%%%%%%%%
A second procedure to obtain the PT kernels emerges from the
perturbative solution of Eq. \eqref{non-hom} under the iterative
process developed in \cite{scoccimarro_98, scoccimarro_01}. We observe
that Eq. \eqref{non-hom} is a system of coupled non-homogeneous, non-linear ordinary differential equations. In EdS the matrix $\Omega$ does not depend on $\eta$ and therefore the solution for the homogeneous problem will be given by:
\begin{equation}
\Psi^{(1)}(\vk, \eta) = \mathrm{e}^{\Omega (\eta-\eta_0)} \Psi^{(0)}(\vk, \eta_0)
\end{equation}
where $\Psi^{(0)}(\vk, \eta_0)$ is the initial condition for the doublet $\Psi$, and 
\begin{equation}
\mathrm{e}^{\Omega (\eta-\eta_0)} = \frac{\mathrm{e}^{\eta-\eta_0}}{5} \left( \begin{array}{cc} 3 & 2 \\ 3 & 2 \end{array} \right) + \frac{\mathrm{e}^{-3/2(\eta-\eta_0)}}{5} \left( \begin{array}{cc} 2 & -2 \\ -3 & 3 \end{array} \right) =: g(\eta, \eta_0)
\end{equation}
is the exponential of the matrix $\Omega$ times the time-lapse, or the Green's function for the homogeneous problem. The solution of the non-homogeneous problem can be written as:

\begin{equation}
\label{Dyson_one}
\Psi_a(\vk, \eta) = g_a^b(\eta, \eta_0) \Psi_b^{(0)}(\vk, \eta_0) +
\int_{\eta_0, \vk_1, \vk_2}^{\eta} \dd\eta' g_a^b(\eta, \eta') 
\gamma_b^{cd} (\vk_1, \vk_2) \Psi_c(\vk_1, \eta') \Psi_d(\vk_2, \eta') \, .
\end{equation}

The solution can be approximated by a Dyson series where, following an iterative strategy, we write $\Psi(\vk, \eta) = \sum_{n=1} \Psi^{(n)}(\vk, \eta)$ and insert this development on the right hand side of Eq. \eqref{Dyson_one}. Terms corresponding to each order can be identified by counting the powers of $\Psi^{(1)}$. Going up to $\Psi^{(3)}$, we have:
\begin{eqnarray}
\label{psi_orders}
\Psi_a(\vk, \eta) & = & \underbrace{g_a^b(\eta, \eta_0) \Psi_b^{(0)}
                        (\vk, \eta_0)}_{\Psi_a^{(1)}(\vk, \eta)} +
                        \underbrace{\int_{\eta_0, \vk_1, \vk_2}^{\eta}
                        \dd \eta' g_a^b(\eta, \eta') \gamma_b^{c d}
                        (\vk_1, \vk_2) \Psi_c^{(1)} (\vk_1, \eta')
                        \Psi_d^{(1)} (\vk_2,
                        \eta')}_{\Psi_a^{(2)}(\vk, \eta)} \nonumber\\
                  & & + \int_{\eta_0, \vk_1, \vk_2, \vk_3}^{\eta} \dd \eta' g_a^b (\eta, \eta') \gamma_b^{cd} (\vk_1, \vk_2 + \vk_3) \Psi_c^{(1)}(\vk_1, \eta') \Psi_d^{(2)}(\vk_2 + \vk_3, \eta') \nonumber\\ & &
+\int_{\eta_0, \vk_1, \vk_2, \vk_3}^{\eta} \dd \eta' g_a^b (\eta, \eta') \gamma_b^{cd} (\vk_1+ \vk_2, \vk_3) \Psi_d^{(1)} (\vk_3, \eta') \Psi_c^{(2)}(\vk_1 + \vk_2, \eta') + \ldots
\end{eqnarray}

Dyson series have natural a diagrammatic representation associated to it, and for Eq. \eqref{psi_orders} we have: 
\begin{fmffile}{three}
\begin{eqnarray}
\label{three_goroff}
\Psi & = & 
\parbox{30mm}{
\begin{fmfgraph}(80,50)
\fmfstraight
\fmfleft{i1}
\fmfright{o1}
\fmf{vanilla,tension=2}{i1,o1}
\end{fmfgraph}}
+ \quad
\parbox{30mm}{
\begin{fmfgraph}(80,50)
\fmfstraight
\fmfleft{i1,i2,i3}
\fmfright{o1,o2}
\fmf{vanilla}{i2,v1,o1}
\fmf{vanilla}{i2,v1,o2}
\end{fmfgraph}}
\nonumber\\ & & + \quad
\parbox{30mm}{
\begin{fmfgraph}(80,50)
\fmfstraight
\fmfleft{i1,i2,i3,i4,i5}
\fmfright{o1,o2,o3,o4,o5}
\fmf{vanilla,tension=2}{i3,v1,o5}
\fmf{phantom,tension=2}{i3,v1,o1}
\fmffreeze
\fmf{phantom,tension=2}{i2,v3,v4,o2}
\fmf{phantom}{v1,v3}
\fmffreeze
\fmf{vanilla,tension=2}{o1,v4,o3}
\fmf{vanilla,tension=2}{v1,v4}
\end{fmfgraph}}
+ \quad 
\parbox{30mm}{
\begin{fmfgraph}(80,50)
\fmfstraight
\fmfleft{i1,i2,i3,i4,i5}
\fmfright{o1,o2,o3,o4,o5}
\fmf{vanilla,tension=2}{i3,v1,o1}
\fmf{phantom,tension=2}{i3,v1,o5}
\fmffreeze
\fmf{phantom,tension=2}{i4,v3,v4,o4}
\fmf{phantom}{v1,v3}
\fmffreeze
\fmf{vanilla,tension=2}{o5,v4,o3}
\fmf{vanilla,tension=2}{v1,v4}
\end{fmfgraph}} + \dots \nonumber\\
\end{eqnarray}
\end{fmffile}
what justifies the meaning of $\gamma$s as vertex couplings.

The solution given in terms of the Dyson series contains not only the
growing modes taken into account in Goroff's method, but also
transient modes \cite{scoccimarro_98}. The PT kernels can be extracted in
this formalism by comparing the terms in the series development of
$\Psi$ with Eqs. \eqref{delta_n} and \eqref{theta_n}. If we collect
only the terms with dominating growing behavior we obtain Goroff's
kernels. The kernels emerging from $\Psi_1$ will be $F$s because they
track couplings of densities and velocities. Each order in
perturbation theory will produce the corresponding $F_n$. From
$\Psi_2$ we collect $G_n$.

We observe that the first integral contributing to $\Psi_1^{(3)}$ in Eq. \eqref{psi_orders} contains the contractions $g_1^1 \gamma_1^{21} \Psi_2^{(0)}  \Psi_1^{(2)} + g_1^2 \gamma_2^{22} \Psi_2^{(0)} \Psi_2^{(2)}$ whereas the second integral brings the contractions $(g_1^1 \gamma_1^{21} \Psi_1^{(0)} + g_1^2 \gamma_2^{22} \Psi_2^{(0)} ) \Psi_2^{(2)}$. Because of the asymmetric coupling, the first two terms in $F_3$ are related to the diagram where moments $\vk_2$ and $\vk_3$ are coupled, and the resultant coupled to $\vk_1$. The third term is related the diagram where $\vk_1$ and $\vk_2$ are coupled first, and then coupled to $\vk_3$.
We will see that this asymmetric coupling structure can be simplified by defining new vertex couplings.

%%%%%%%%%%%%%%%%%%%%%%%%%%%%%%%%%%
\section{Symmetrization procedures}
%%%%%%%%%%%%%%%%%%%%%%%%%%%%%%%%%%

We observe that in Eqs. \eqref{delta_n}, \eqref{theta_n} we integrate $F_n(\vk_1, \ldots, \vk_n)$ and $G_n(\vk_1, \ldots, \vk_n)$ multiplied by a combination of functions symmetric under  permutations of $\vk_1, \ldots, \vk_n$, and therefore only the symmetric part of the kernels give a non-vanishing contribution to $\delta^{(n)}$ and $\theta^{(n)}$. The kernels to be considered in Perturbation Theory are, therefore, the full symmetrized versions of $F_n$ and $G_n$, defined as:

\begin{equation}
\label{f_SYM}
F_n^{SYM} (\vk_1, \ldots, \vk_n) = \frac{1}{n!} \sum_{\vq_1, \ldots, \vq_n \in \pi(\vk_1, \ldots, \vk_n)} F_n(\vq_1, \ldots, \vq_n) \, ,
\end{equation}
\begin{equation}
\label{g_SYM}
G_n^{SYM} (\vk_1, \ldots, \vk_n) = \frac{1}{n!} \sum_{\vq_1, \ldots, \vq_n \in \pi(\vk_1, \ldots, \vk_n)} G_n(\vq_1, \ldots, \vq_n) \, ,
\end{equation}
where $\pi(\vk_1, \ldots, \vk_n)$ denotes the set of permutations of the symbols $\vk_1, \ldots, \vk_n$. Both Goroff's and Scoccimarro's methods will produce the same symmetrized kernels. One further step towards $F_n^{SYM}, G_n^{SYM}$ can be made in Scoccimarro's method by promoting the vertex couplings $\gamma$ to the \emph{symmetric vertex coupling} $\gamma^s$, introduced in \cite{RPT}, and defined as:
\begin{equation}
\gamma_1^{s21}(\vk_1, \vk_2) := \delta_D(\vk - \vk_1 - \vk_2)  \frac{\alpha(\vk_1, \vk_2)}{2}   \qquad
\gamma_1^{s12}(\vk_1, \vk_2) := \delta_D(\vk - \vk_1 - \vk_2)  \frac{\alpha(\vk_2, \vk_1)}{2}
\end{equation}
\begin{equation}
\gamma_2^{s22}(\vk_1, \vk_2) = \gamma_2^{22}(\vk_1, \vk_2) = \delta_D(\vk - \vk_1 - \vk_2)  \beta(\vk_1, \vk_2)  \, .
\end{equation}
Looking at the right hand side of Eq. \eqref{non-hom}, we clearly see that the promotion $\gamma \to \gamma^s$ makes sense
because we can perform a change of variables when integrating over the momenta $\vk_1, \vk_2$. This shows \emph{why} it is reasonable to introduce a symmetric vertex, i.e., solving Eq. \eqref{non-hom} with $\gamma \to \gamma^s$ should conduct to the same results when we look at the full-symmetrized version of the kernels. It does not clarify, however, \emph{how} the equivalence of the solutions is produced at the level of the kernels.

One could naively expect that the kernels obtained with symmetric
vertex would be the ones obtained by inserting symmetrized kernels on
the recurrence relations \eqref{F_recurrence},
\eqref{G_recurrence}. The replacement $\gamma \to \gamma^s$ has, as we
shall see, a more subtle action. In order to describe the equivalence of the kernels produced by both methods, we will look at the kernels order by order. 

%%%%%%%%%%%%%%%%%%%%%%%%%%%%%%
\subsection{Equivalence at second order}
%%%%%%%%%%%%%%%%%%%%%%%%%%%%%%
The first non-trivial kernels appears in PT at second order calculations. From the recurrence relations \eqref{F_recurrence}, \eqref{G_recurrence} we have:

\begin{equation}
\label{f2_g}
F_2(\vk_{1}, \vk_{2})  =  \frac{1}{7} \left(5 \alpha(\vk_{1}, \vk_{2}) + 2 \beta (\vk_{1}, \vk_{2}) \right) \, ,\qquad \qquad
G_2(\vk_{1}, \vk_{2})  =  \frac{1}{7} \left(3 \alpha(\vk_{1}, \vk_{2}) + 4 \beta (\vk_{1}, \vk_{2}) \right) \, .
\end{equation}
After symmetrization under permutation of momenta:
\begin{eqnarray}
\label{f2_gs}
F_2^{SYM}(\vk_{1}, \vk_{2}) & = & \frac{1}{7} \left[5 \left[\frac{\alpha(\vk_{1}, \vk_{2}) + \alpha(\vk_{2}, \vk_{1})}{2} \right] + 2 \beta (\vk_{1}, \vk_{2}) \right] \, ,\\
G_2^{SYM}(\vk_{1}, \vk_{2}) & = & \frac{1}{7} \left[3 \left[ \frac{\alpha(\vk_{1}, \vk_{2}) + \alpha(\vk_{2}, \vk_{1})}{2} \right]+ 4 \beta (\vk_{1}, \vk_{2}) \right] \, .
\end{eqnarray}
This output follows directly from the second order term in \eqref{psi_orders} after the replacement $\gamma \to \gamma^s$. Symmetrization by summing over momenta permutations or the replacement $\gamma \to \gamma^s$ produce, therefore, directly the same output at the lowest non-trivial order.

%%%%%%%%%%%%%%%%%%%%%%%%%%%%%%
\subsection{Equivalence at third order}
%%%%%%%%%%%%%%%%%%%%%%%%%%%%%%

This is the first order where the symmetrization properties of the
kernels lead to interesting results, what motivates us to  discuss in
detail the properties of $F_3^{SYM}, G_3^{SYM}$.

\subsubsection{Goroff's method}
Following the recurrence relations \eqref{F_recurrence}, \eqref{G_recurrence}, we obtain for $n=3$:

\begin{eqnarray}
\label{f3_g}
F_3(\vk_1, \vk_2, \vk_3) & = & \frac{1}{18} \Big[ 7 \alpha(\vk_1, \vk_2+\vk_3) F_2(\vk_2, \vk_3)  + 2 \beta(\vk_1, \vk_2 + \vk_3) G_2(\vk_2, \vk_3)   \nonumber\\ & & + 
G_2(\vk_1, \vk_2) \left[ 7 \alpha(\vk_1 + \vk_2, \vk_3) + 2 \beta(\vk_1 + \vk_2, \vk_3) \right] \Big] 
\end{eqnarray}
\begin{eqnarray}
\label{g3_g}
G_3(\vk_1, \vk_2, \vk_3) & = & \frac{1}{18} \Big[ 3 \alpha(\vk_1, \vk_2+\vk_3) F_2(\vk_2, \vk_3)  + 6 \beta(\vk_1, \vk_2 + \vk_3) G_2(\vk_2, \vk_3)   \nonumber\\ & & + 
G_2(\vk_1, \vk_2) \left[ 3 \alpha(\vk_1 + \vk_2, \vk_3) + 6 \beta(\vk_1 + \vk_2, \vk_3) \right] \Big] \, .
\end{eqnarray}
The first two terms in $F_3, G_3$ are related to the diagram where moments $\vk_2$ and $\vk_3$ are coupled on a first vertex and their resultant is coupled to $\vk_1$ on a second vertex. The third term corresponds to the diagram where $\vk_1$ and $\vk_2$ are coupled first, and then coupled to $\vk_3$. The coupling structure is apprehended from Eqs. \eqref{f3_g}, \eqref{g3_g} by looking at which momenta are summed on the arguments of $\alpha$ and $\beta$. The $F_2$s and $G_2$s have each of the momenta in the sum as arguments.
In order to shorten the notation we will only write the indexes of the momenta in the arguments of $\alpha F_2$, $\alpha G_2$, $\beta G_2$, separating the terms appearing on each of the arguments of $\alpha$ or $\beta$ by parentheses. For $F_3$, for example, we will make the following associations:

\begin{fmffile}{threef3}
\begin{center}
\begin{equation}
\label{three_goroff}
\frac{1}{18} \Big[ 7 \alpha(\vk_1, \vk_2+\vk_3) F_2(\vk_2, \vk_3)  + 2 \beta(\vk_1, \vk_2 + \vk_3) G_2(\vk_2, \vk_3) \Big] \to
\parbox{40mm}{
\begin{fmfgraph}(80,30)
\fmfstraight
\fmfleft{i1,i2,i3,i4,i5}
\fmfright{o1,o2,o3,o4,o5}
\fmf{vanilla,tension=2}{i3,v1,o5}
\fmf{phantom,tension=2}{i3,v1,o1}
\fmffreeze
\fmf{phantom,tension=2}{i2,v3,v4,o2}
\fmf{phantom}{v1,v3}
\fmffreeze
\fmf{vanilla,tension=2}{o1,v4,o3}
\fmf{vanilla,tension=2}{v1,v4}
\end{fmfgraph}} \to (1)(23) \nonumber
\end{equation}
\begin{equation}
\frac{1}{18} \Big[ G_2(\vk_1, \vk_2) \big( 7 \alpha(\vk_1 + \vk_2, \vk_3) + 2 \beta(\vk_1 + \vk_2, \vk_3) \big) \Big] \to
\parbox{40mm}{
\begin{fmfgraph}(80,30)
\fmfstraight
\fmfleft{i1,i2,i3,i4,i5}
\fmfright{o1,o2,o3,o4,o5}
\fmf{vanilla,tension=2}{i3,v1,o1}
\fmf{phantom,tension=2}{i3,v1,o5}
\fmffreeze
\fmf{phantom,tension=2}{i4,v3,v4,o4}
\fmf{phantom}{v1,v3}
\fmffreeze
\fmf{vanilla,tension=2}{o5,v4,o3}
\fmf{vanilla,tension=2}{v1,v4}
\end{fmfgraph}}  \to (12)(3) \nonumber
\end{equation}
\end{center}
\end{fmffile}

We will call $(1)(23)$ and $(12)(3)$ \emph{repartitions}. The recurrence relations give numerical factors appearing on each of the kernels, what allows us to reconstruct the kernels if we know the repartitions. The same simplified notation can be used for higher orders, both for $F$ and $G$, though the numerical factors will be different for each kernel at each order. Using this notation, we can write
\begin{eqnarray}
\label{f3_perm}
F_3^{SYM}(\vk_1, \vk_2, \vk_3) & = & \frac{1}{6} \Big[ (1)(23) + (12)(3) + (2)(31) + (23)(1) + (3)(12) + (31)(2) \nonumber\\ & & + (1)(32) + (13)(2) + (2)(13) + (21)(3) + (3)(21) + (32)(1) \Big] \nonumber\\ & = &
\frac{1}{3} \Big[ (1)(23)^{\pi} + (2)(13)^{\pi} + (3)(12)^{\pi} +
                                                                                                                                                                                                                     (12)^{\pi}(3)
                                                                                                                                                                                                                     +
                                                                                                                                                                                                                     (13)^{\pi}(2)
                                                                                                                                                                                                                     +
                                                                                                                                                                                                                     (23)^{\pi}(1)
                                                                                                                                                                                                                     \Big]
                                                                                                                                                                                                                     \, .
\end{eqnarray}
On the first two lines we wrote the six permutations of $\{\vk_1,
\vk_2, \vk_3\}$ in the repartitions that correspond to the structure
of the kernel $F_3$. On the last line we defined $( \, \, \, \,
)^{\pi}$ to denote the sum over permutations of $n$ elements divided
by the number of permutations. Explicitly $(1)(23)^{\pi} = \frac{1}{2}
\big( (1)(23) + (1)(32) \big)$. We can associate pairs of terms being summed in Eq. \eqref{f3_perm} in two different ways, that can be graphically represented as:

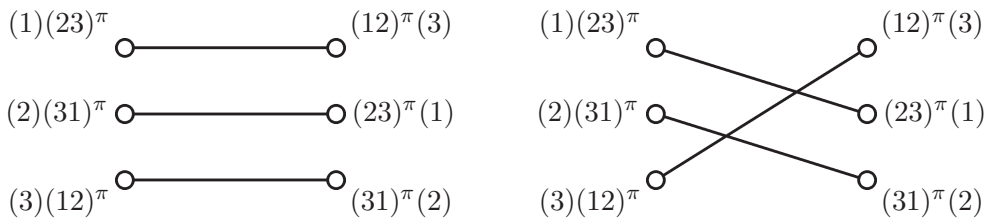
\begin{fmffile}{three_connexions}
\begin{equation}
\parbox{40mm}{
\begin{fmfgraph*}(80,50)
\fmfstraight
\fmfleft{i1,i2,i3}
\fmfright{o1,o2,o3}
\fmfv{label=$(3)(12)^\pi$,decor.shape=circle,decor.filled=empty,decor.size=.08w}{i1}
\fmfv{label=$(2)(31)^\pi$,decor.shape=circle,decor.filled=empty,decor.size=.08w}{i2}
\fmfv{label=$(1)(23)^\pi$,decor.shape=circle,decor.filled=empty,decor.size=.08w}{i3}
\fmfv{label=$(31)^\pi(2)$,decor.shape=circle,decor.filled=empty,decor.size=.08w}{o1}
\fmfv{label=$(23)^\pi(1)$,decor.shape=circle,decor.filled=empty,decor.size=.08w}{o2}
\fmfv{label=$(12)^\pi(3)$,decor.shape=circle,decor.filled=empty,decor.size=.08w}{o3}
\fmf{vanilla,tension=2}{i1,o1}
\fmf{vanilla,tension=2}{i2,o2}
\fmf{vanilla,tension=2}{i3,o3}
\end{fmfgraph*}} \qquad \qquad \qquad \qquad
\parbox{40mm}{
\begin{fmfgraph*}(80,50)
\fmfstraight
\fmfleft{i1,i2,i3}
\fmfright{o1,o2,o3}
\fmfv{label=$(3)(12)^\pi$,decor.shape=circle,decor.filled=empty,decor.size=.08w}{i1}
\fmfv{label=$(2)(31)^\pi$,decor.shape=circle,decor.filled=empty,decor.size=.08w}{i2}
\fmfv{label=$(1)(23)^\pi$,decor.shape=circle,decor.filled=empty,decor.size=.08w}{i3}
\fmfv{label=$(31)^\pi(2)$,decor.shape=circle,decor.filled=empty,decor.size=.08w}{o1}
\fmfv{label=$(23)^\pi(1)$,decor.shape=circle,decor.filled=empty,decor.size=.08w}{o2}
\fmfv{label=$(12)^\pi(3)$,decor.shape=circle,decor.filled=empty,decor.size=.08w}{o3}
\fmf{vanilla,tension=2}{i3,o2}
\fmf{vanilla,tension=2}{i2,o1}
\fmf{vanilla,tension=2}{i1,o3}
\end{fmfgraph*}} 
\nonumber
\end{equation}
\end{fmffile}

The first arrangement -- corresponding to the parallel association -- yields:

\begin{eqnarray}
\label{F3_SYM_s}
F_3^{SYM}(\vk_1, \vk_2, \vk_3) & = & \frac{1}{3} \Big[ \big( (1)(23)^{\pi} + (12)^{\pi}(3) \big) + \big( (2)(13)^{\pi} + (23)^{\pi}(1) \big) + \big( (3)(12)^{\pi} + (13)^{\pi}(2) \big)  \Big] \nonumber\\ & =: & \frac{1}{3} \Big[ F^s_3(\vk_1, \vk_2, \vk_3) + F^s_3(\vk_2, \vk_3, \vk_1) + F^s_3(\vk_3, \vk_1, \vk_2) \Big] \, ,
\end{eqnarray}
where 
\begin{eqnarray}
F_3^s(\vk_1, \vk_2, \vk_3) & := & \frac{1}{18} \Big[ 7 \alpha(\vk_1, \vk_2+\vk_3) F^{SYM}_2(\vk_2, \vk_3)  + 2 \beta(\vk_1, \vk_2 + \vk_3) G^{SYM}_2(\vk_2, \vk_3)  \nonumber\\ & & + 
G^{SYM}_2(\vk_1, \vk_2) \left[ 7 \alpha(\vk_1 + \vk_2, \vk_3) + 2 \beta(\vk_1 + \vk_2, \vk_3) \right] \Big] \, .
\end{eqnarray}
This shows that we can generate $F_3^{SYM}$ by inserting $F_2^{SYM}$ and $G_2^{SYM}$ on the recurrence relation \eqref{f3_g} and summing over the cyclic permutations of $\{\vk_1, \vk_2, \vk_3\}$. We observe, however, that the symmetries properties of $F_3^s$ under permutation of momenta are not superior to those of $F_3$, i.e., in terms of its structure, we can separate $F_3^s$ schematically as:
\begin{equation}
\label{F3_schem}
F_3^s(\vk_1, \vk_2, \vk_3) = T_1 + \frac{T_2}{|\vk_2 + \vk_3|^2} + \frac{T_3}{|\vk_1 + \vk_2|^2} \, ,
\end{equation}
where  $T_1$, $T_2$, $T_3$ are polynomial functions of $k_1^2, k_2^2, k_3^3, \vk_1 \cdot \vk_2, \vk_1 \cdot \vk_3$, and $\vk_2 \cdot \vk_3$. We observe, hence, that $F_3^s(\vk_1, \vk_2, \vk_3)$ has dependence on the angles between $\vk_2$ and $\vk_3$, and $\vk_1$ and $\vk_2$ on the denominators.

To the second way of organising the terms -- the crossed link -- corresponds a more symmetrical output:

\begin{eqnarray}
\label{F3_SYM_s}
F_3^{SYM}(\vk_1, \vk_2, \vk_3) & = & \frac{1}{3} \Big[ \big( (1)(23)^{\pi} + (23)^{\pi}(1) \big) + \big( (2)(13)^{\pi} + (13)^{\pi}(2) \big) + \big( (3)(12)^{\pi} + (12)^{\pi}(3) \big)  \Big] \nonumber\\ & =: & \frac{1}{3} \Big[ \tilde{F}_3(\vk_1, \vk_2, \vk_3) + \tilde{F}_3(\vk_2, \vk_3, \vk_1) + \tilde{F}_3(\vk_3, \vk_1, \vk_2) \Big] \, 
\end{eqnarray}
with
\begin{eqnarray}
\label{F3_tilde}
\tilde{F}_3(\vk_1, \vk_2, \vk_3) & := & \frac{1}{18} \Big[ 7 \alpha(\vk_1, \vk_2 + \vk_3) F_2^{SYM}(\vk_2, \vk_3) + 2 \beta(\vk_1, \vk_2 + \vk_3) G_2^{SYM}(\vk_2, \vk_3)   \nonumber\\ & &  +  G_2^{SYM}(\vk_2, \vk_3) \left( 7 \alpha(\vk_2 + \vk_3, \vk_1) + 2 \beta(\vk_2 + \vk_3, \vk_1) \right) \Big] \, .
\end{eqnarray}

The nice feature of $\tilde{F}_3$ is its symmetry under the exchange
of its two last arguments. The analogous version of Eq. \eqref{F3_schem} for $\tilde{F}_3$ is: 
\begin{equation}
\label{F3_tilde_schem}
\tilde{F}_3(\vk_1, \vk_2, \vk_3) = T_1' + \frac{ T_2'}{|\vk_2 + \vk_3|^2} \, .
\end{equation}
As before, $T'_1$, $T'_2$ are polynomial functions of $k_1^2, k_2^2, k_3^3, \vk_1 \cdot \vk_2, \vk_1 \cdot \vk_3$, and $\vk_2 \cdot \vk_3$. $\tilde{F}_3$ has, therefore, a simpler structure than $F_3^s$ because only the angle between $\vk_2$ and $\vk_3$ appears on the denominator.

In terms of the mode coupling structure, $\tilde{F}_3$ would
correspond to a diagram with structure:
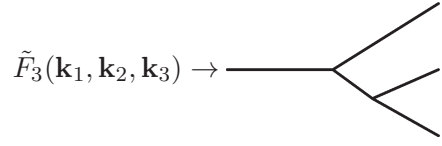
\begin{fmffile}{threef3tilde}
\begin{equation}
\tilde{F}_3(\vk_1, \vk_2, \vk_3) \to
\parbox{40mm}{
\begin{fmfgraph}(80,50)
\fmfstraight
\fmfleft{i1,i2,i3,i4,i5}
\fmfright{o1,o2,o3,o4,o5}
\fmf{vanilla,tension=2}{i3,v1,o5}
\fmf{phantom,tension=2}{i3,v1,o1}
\fmffreeze
\fmf{phantom,tension=2}{i2,v3,v4,o2}
\fmf{phantom}{v1,v3}
\fmffreeze
\fmf{vanilla,tension=2}{o1,v4,o3}
\fmf{vanilla,tension=2}{v1,v4}
\end{fmfgraph}} \nonumber
\end{equation}
\end{fmffile}

The same structure is satisfied by $G_3$. For instance,
\begin{equation}
\label{F3_SYM_curly}
G_3^{SYM}(\vk_1, \vk_2, \vk_3) = \frac{1}{3} \Big[ \tilde{G}_3(\vk_1, \vk_2, \vk_3) + \tilde{G}_3(\vk_2, \vk_3, \vk_1) + \tilde{G}_3(\vk_3, \vk_1, \vk_2) \Big] \, 
\end{equation}
with
\begin{eqnarray}
\label{G3_tilde}
\tilde{G}_3(\vk_1, \vk_2, \vk_3) & = & \frac{1}{18} \Big[ 3
                                       \alpha(\vk_1, \vk_2 + \vk_3)
                                       F_2^{SYM}(\vk_2, \vk_3) + 6
                                       \beta(\vk_1, \vk_2 + \vk_3)
                                       G_2^{SYM}(\vk_2, \vk_3)
                                       \nonumber\\ & &  +
                                                       G_2^{SYM}(\vk_2,
                                                       \vk_3)  \left( 3 \alpha(\vk_2 + \vk_3, \vk_1) + 6 \beta(\vk_2 + \vk_3, \vk_1) \right) \Big] \, .
\end{eqnarray}
%%%%%%%%%%%%%%%%%%%%%%%%%%%%
\subsubsection{Scoccimarro's method}
%%%%%%%%%%%%%%%%%%%%%%%%%%%%
If we proceed the iterative solution of Eq. \eqref{non-hom} to include third order terms -- as shown in Eq. \eqref{psi_orders} -- with the non-symmetric vertex, the kernels shown in Eqs. \eqref{f3_g} and \eqref{g3_g} are directly produced. If we use the symmetric vertex $\gamma^s$, on the other hand, the iterative evolution of the growing modes will produce $\tilde{F}_3, \tilde{G}_3$ as outputs. This is why the substitution $\gamma \to \gamma^s$ is subtle: $\tilde{F}_3$ has to be understood as building blocks of $F_3^{SYM}$, but constructed by smartly grouping terms in Eq. \eqref{f3_perm}. The same holds for $G_3^{SYM}$.

The possibility of writing $F_3^{SYM}$ in terms of the cyclic permutations of $\tilde{F}_3$ allows to rewrite Eq. \eqref{psi_orders} as
\begin{equation}
\label{psi_orders_sym}
\Psi_a(\vk, \eta)  =   \Psi_b^{(1)} (\vk, \eta) + \Psi_d^{(2)} (\vk_2, \eta) + 2 \int_{\eta_0}^{\eta} d \eta' g_a^b (\eta, \eta') \gamma_b^{cd} (\vk_1, \vk_2 + \vk_3) \Psi_c^{(1)}(\vk_1, \eta') \Psi_d^{(2)}(\vk_2 + \vk_3, \eta') + \ldots
\end{equation}
where the symmetry factor $2$ appears in front of the integral, corresponding to the fact that the two different ways of writing the diagram of three external legs are equivalent after symmetrization.

As conclusion, $F_3^{SYM}$ obtained from the Scoccimarro's method with symmetrized vertex corresponds to the fully symmetric combination of the $F_3$ obtained from the Goroff's method, but the equivalence is only explicit after the introduction of the tilded kernel $\tilde{F}_3$.

%%%%%%%%%%%%%%%%%%%%%%%%%%%
\subsection{Structure at fourth order}
%%%%%%%%%%%%%%%%%%%%%%%%%%%

We will concentrate here on $F_4$ because $G_4$ has the same properties. We observe first that, from the recurrence relations, we obtain an expression for $F_4$ that can be decomposed in two terms according to their structure:

\begin{equation}
F_4(\vk_1, \vk_2, \vk_3, \vk_4) = F_4^1(\vk_1, \vk_2, \vk_3, \vk_4) + F_4^2(\vk_1, \vk_2, \vk_3, \vk_4)
\end{equation}
where the \emph{subkernels} $F_4^{1, 2}$ are given by:
\begin{eqnarray}
F_4^1(\vk_1, \vk_2, \vk_3, \vk_4) & = & \frac{1}{33} \Big[ \big[ 9 \alpha(\vk_1, \vk_2 + \vk_2 + \vk_3) F_3(\vk_2, \vk_3, \vk_4)  \nonumber\\ & & \hspace{0.6cm} + 2 \beta(\vk_1, \vk_2 + \vk_3 + \vk_4) G_3(\vk_2, \vk_3, \vk_4) \big]  \nonumber\\ & & \hspace{0.6cm} + G_3(\vk_1, \vk_2, \vk_3) \big[ 9 \alpha(\vk_1 + \vk_2 + \vk_3, \vk_4) \nonumber\\ & & \hspace{0.6cm}+ 2 \beta(\vk_1 + \vk_2 + \vk_3, \vk_4) \big] \Big] \, ,
\end{eqnarray}
and
\begin{eqnarray}
F_4^2(\vk_1, \vk_2, \vk_3, \vk_4) & = &  \frac{1}{33} \Big[ G_2(\vk_1, \vk_2) \big[ 9 \alpha(\vk_1 + \vk_2, \vk_3 + \vk_4) F_2(\vk_3, \vk_4) \nonumber\\ & & \hspace{0.6cm} + 2 \beta(\vk_1 + \vk_2, \vk_3 + \vk_4) G_2(\vk_3, \vk_4) \big] \Big] \, .
\end{eqnarray}

In order to describe the 24 permutations of $\{ \vk_1, \vk_2, \vk_3, \vk_4 \}$ in $F_4^1$, we can look at four sets of six components, where the first momenta is fixed and the six permutations of the three other elements is performed. Considering the six permutations of $\{\vk_2, \vk_3, \vk_4\}$ keeping $\vk_1$ fixed on the term $\alpha(\vk_1, \vk_2 + \vk_3 + \vk_4) F_3 (\vk_2, \vk_3, \vk_4)$, for example, we have $\alpha(\vk_1, \vk_2+\vk_3+\vk_4) F_3^{SYM}(\vk_2, \vk_3, \vk_4)$. The same holds for the other terms, and reintroducing the notation of repartitions, we can write:
\begin{eqnarray}
F_4^{1, SYM}(\vk_1, \vk_2, \vk_3, \vk_4) & = & \frac{1}{4} \Big[ (1)(234)^{\pi} + (2)(341)^{\pi} + (3)(412)^{\pi} + (4)(123)^{\pi} \nonumber\\ & & \hspace{0.5cm}+ (123)^{\pi}(4) + (234)^{\pi}(1) + (341)^{\pi}(2) + (412)^{\pi}(3) \Big] \, .
\end{eqnarray}
Here there are also two possible ways of arranging the terms:

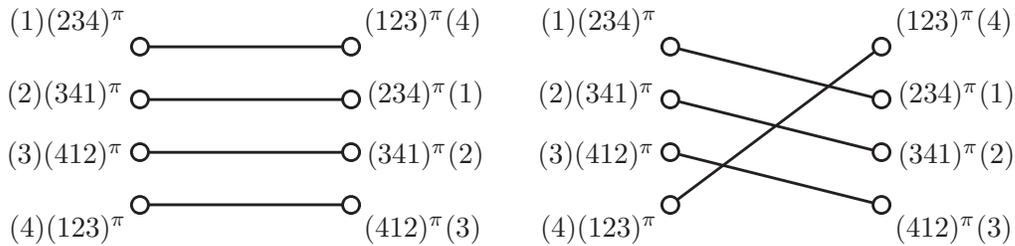
\begin{fmffile}{four_connexions}
\begin{equation}
\parbox{40mm}{
\begin{fmfgraph*}(80,60)
\fmfstraight
\fmfleft{i1,i2,i3,i4}
\fmfright{o1,o2,o3,o4}
\fmfv{label=$(4)(123)^\pi$,lable.angle=0,decor.shape=circle,decor.filled=empty,decor.size=.08w}{i1}
\fmfv{label=$(3)(412)^\pi$,lable.angle=0,decor.shape=circle,decor.filled=empty,decor.size=.08w}{i2}
\fmfv{label=$(2)(341)^\pi$,lable.angle=0,decor.shape=circle,decor.filled=empty,decor.size=.08w}{i3}
\fmfv{label=$(1)(234)^\pi$,lable.angle=0,decor.shape=circle,decor.filled=empty,decor.size=.08w}{i4}
\fmfv{label=$(412)^\pi(3)$,lable.angle=0,decor.shape=circle,decor.filled=empty,decor.size=.08w}{o1}
\fmfv{label=$(341)^\pi(2)$,lable.angle=0,decor.shape=circle,decor.filled=empty,decor.size=.08w}{o2}
\fmfv{label=$(234)^\pi(1)$,lable.angle=0,decor.shape=circle,decor.filled=empty,decor.size=.08w}{o3}
\fmfv{label=$(123)^\pi(4)$,lable.angle=0,decor.shape=circle,decor.filled=empty,decor.size=.08w}{o4}
\fmf{vanilla,tension=2}{i1,o1}
\fmf{vanilla,tension=2}{i2,o2}
\fmf{vanilla,tension=2}{i3,o3}
\fmf{vanilla,tension=2}{i4,o4}
\end{fmfgraph*}} \qquad \qquad \qquad \qquad 
\parbox{40mm}{
\begin{fmfgraph*}(80,60)
\fmfstraight
\fmfleft{i1,i2,i3,i4}
\fmfright{o1,o2,o3,o4}
\fmfv{label=$(4)(123)^\pi$,decor.shape=circle,decor.filled=empty,decor.size=.08w}{i1}
\fmfv{label=$(3)(412)^\pi$,decor.shape=circle,decor.filled=empty,decor.size=.08w}{i2}
\fmfv{label=$(2)(341)^\pi$,decor.shape=circle,decor.filled=empty,decor.size=.08w}{i3}
\fmfv{label=$(1)(234)^\pi$,decor.shape=circle,decor.filled=empty,decor.size=.08w}{i4}
\fmfv{label=$(412)^\pi(3)$,decor.shape=circle,decor.filled=empty,decor.size=.08w}{o1}
\fmfv{label=$(341)^\pi(2)$,decor.shape=circle,decor.filled=empty,decor.size=.08w}{o2}
\fmfv{label=$(234)^\pi(1)$,decor.shape=circle,decor.filled=empty,decor.size=.08w}{o3}
\fmfv{label=$(123)^\pi(4)$,decor.shape=circle,decor.filled=empty,decor.size=.08w}{o4}
\fmf{vanilla,tension=2}{i4,o3}
\fmf{vanilla,tension=2}{i3,o2}
\fmf{vanilla,tension=2}{i2,o1}
\fmf{vanilla,tension=2}{i1,o4}
\end{fmfgraph*}} 
\nonumber
\end{equation}
\end{fmffile}

For the parallel link case, we obtain
\begin{eqnarray}
F_4^{1, SYM}(\vk_1, \vk_2, \vk_3, \vk_4) & = & \frac{1}{4} \Big[ \big( (1)(234)^{\pi} + (123)^{\pi}(4) \big) + \big( (2)(314)^{\pi} + (234)^{\pi}(1) \big) \nonumber\\ & & \hspace{0.5cm} + \big( (3)(412)^{\pi} + (341)^{\pi}(2) \big) + \big( (4)(123)^{\pi} + (412)^{\pi}(3) \big) \Big] \nonumber\\ & =: &
\frac{1}{4} \Big[ F_4^{1,s}(\vk_1, \vk_2, \vk_3, \vk_4) + F_4^{1,s}(\vk_2, \vk_3, \vk_4, \vk_1) \nonumber\\ & & \hspace{0.5cm} + F_4^{1,s}(\vk_3, \vk_4, \vk_1, \vk_2) + F_4^{1,s}(\vk_4, \vk_1, \vk_2, \vk_3) \Big] \, ,
\end{eqnarray}
i.e., $F_4^{1,s}$ is generated by the recurrence relation \eqref{F_recurrence} by inserting $F_3^{SYM}, G_3^{SYM}$ instead of $F_3, G_3$. What interest us more, however, is the output from the crossed link arrangement of terms:
\begin{eqnarray}
F_4^{1, SYM}(\vk_1, \vk_2, \vk_3, \vk_4) & = & \frac{1}{4} \Big[ \big( (1)(234)^{\pi} + (234)^{\pi}(1) \big) + \big( (2)(341)^{\pi} + (341)^{\pi}(2) \big) \nonumber\\ & & \hspace{0.6cm} + \big( (3)(412)^{\pi} + (412)^{\pi}(3) \big) + \big( (4)(123)^{\pi} + (123)^{\pi}(4) \big) \Big] \nonumber\\ & =: &
\frac{1}{4} \Big[ \tilde{F}_4^{1}(\vk_1, \vk_2, \vk_3, \vk_4) + \tilde{F}_4^{1}(\vk_2, \vk_3, \vk_4, \vk_1) \nonumber\\& & \hspace{0.5cm}+ \tilde{F}_4^{1}(\vk_3, \vk_4, \vk_1, \vk_2) + \tilde{F}_4^{1}(\vk_4, \vk_1, \vk_2, \vk_3) \Big] \, .
\end{eqnarray}
Explicitly $\tilde{F}_4^1$ is given by:
\begin{eqnarray}
\label{F4_a_tilde}
\tilde{F}_4^1(\vk_1, \vk_2, \vk_3, \vk_4) & = & \frac{1}{33} \Big[ \big[9 \alpha(\vk_1, \vk_2 + \vk_2 + \vk_3) F_3^{SYM}(\vk_2, \vk_3, \vk_4) \nonumber\\ & & \hspace{0.6cm} + 2 \beta(\vk_1, \vk_2 + \vk_3 + \vk_4) G_3^{SYM}(\vk_2, \vk_3, \vk_4) \big] \nonumber\\ & & \hspace{0.6cm} + G_3^{SYM}(\vk_2, \vk_3, \vk_4) \big[ 9 \alpha(\vk_2 + \vk_3 + \vk_4, \vk_1) \nonumber\\ & & \hspace{0.6cm} + 2 \beta(\vk_2 + \vk_3 + \vk_4, \vk_1) \big] \Big] \, ,
\end{eqnarray}
that is symmetric under the exchange of its three last arguments. The iterative solution with symmetrized coupling constants leads to $\tilde{F}_4^{1}$. 

We observe that $F_4^{1,s}(\vk_1, \vk_2, \vk_3, \vk_4)$ has dependence on the angles between $\vk_1$ and all the other vectors $\vk_2, \vk_3, \vk_4$ on the denominator, whereas $\tilde{F}_4^1(\vk_1, \vk_2, \vk_3, \vk_4)$ has only dependence on $\vk_2 \cdot \vk_3, \vk_2 \cdot \vk_4$, and $\vk_3 \cdot \vk_4$ on the denominators. The possibility of separating the dependence on one of the vectors on the denominator of $\tilde{F}_4^1$ can provide simpler integration routines for angular variables.

We turn now our attention to the second subkernel $F_4^2$. Summing over permutations we obtain:

\begin{eqnarray}
F_4^{2, SYM} (\vk_1, \vk_2, \vk_3, \vk_4) &  = & \frac{1}{6} \Big[ (12)^{\pi}(34)^{\pi} + (13)^{\pi}(24)^{\pi} + (14)^{\pi}(23)^{\pi} \nonumber\\ & & \hspace{0.5cm} + (23)^{\pi}(14)^{\pi} + (24)^{\pi}(13)^{\pi} + (34)^{\pi}(12)^{\pi} \Big] \nonumber\\ & =:&
\frac{1}{6} \left[  F_4^{2,s} (\vk_1, \vk_2, \vk_3, \vk_4)  + F_4^{2,s} (\vk_1, \vk_3, \vk_2, \vk_4) + F_4^{2,s} (\vk_1, \vk_4, \vk_2, \vk_3) \right. \nonumber\\ & & \left. + F_4^{2,s} (\vk_2, \vk_3, \vk_1, \vk_4) + F_4^{2,s} (\vk_2, \vk_4, \vk_1, \vk_3) + F_4^{2,s} (\vk_3, \vk_4, \vk_1, \vk_2) \right] \nonumber \, .
\end{eqnarray}
As before, $F_4^{2,s}$ is generated by inserting $F_2^{SYM}, G_2^{SYM}$ in the recurrence relations:
\begin{eqnarray}
F_4^{2,s}(\vk_1, \vk_2, \vk_3, \vk_4) & = &  \frac{1}{33} \Big[ G_2^{SYM}(\vk_1, \vk_2) \big[ 9 \alpha(\vk_1 + \vk_2, \vk_3 + \vk_4) F_2^{SYM}(\vk_3, \vk_4) \nonumber\\ & & \hspace{0.6cm} + 2 \beta(\vk_1 + \vk_2, \vk_3 + \vk_4) G_2^{SYM}(\vk_3, \vk_4) \big] \Big] \, .
\end{eqnarray}

Since the recurrence relations generate only one family of repartitions with structure $(12)(34)$, there is no alternative way of grouping terms, and therefore we will define $\tilde{F}_4^2=F_4^{2,s}$.

Diagrammatically we have the correspondences:

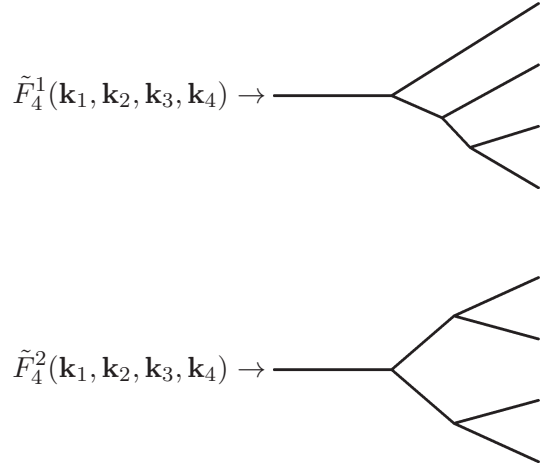
\begin{fmffile}{fourdown}
\begin{equation}
\tilde{F}_4^1(\vk_1, \vk_2, \vk_3, \vk_4) \to
\parbox{40mm}{
\begin{fmfgraph}(100,70)
\fmfstraight
\fmfleft{i1,i2,i3,i4,i5,i6,i7}
\fmfright{o1,o2,o3,o4,o5,o6,o7}
\fmf{vanilla,tension=2}{i4,v1,o7}
\fmf{phantom,tension=2}{i4,v1,o1}
\fmf{vanilla}{i4,v1}
\fmffreeze
\fmf{phantom,tension=2}{i2,v2,v3,v4,o2}
\fmf{phantom,tension=2}{i3,v5,v6,o3}
\fmf{phantom}{v1,v5}
\fmf{vanilla}{v1,v6}
\fmf{phantom}{v1,v2}
\fmf{phantom}{v5,v3}
\fmffreeze
\fmf{vanilla,tension=2}{o1,v4,o3}
\fmf{vanilla,tension=2}{v6,v4}
\fmf{vanilla,tension=2}{v6,o5}
\end{fmfgraph}} \nonumber
\end{equation}
\end{fmffile}

\begin{fmffile}{foursym}
\begin{equation}
\tilde{F}_4^{2}(\vk_1, \vk_2, \vk_3, \vk_4) \to
\parbox{40mm}{
\begin{fmfgraph}(100,70)
\fmfstraight
\fmfleft{i1,i2,i3,i4,i5,i6,i7}
\fmfright{o1,o2,o3,o4,o5,o6,o7}
\fmf{phantom,tension=2}{i4,v1,o7}
\fmf{phantom,tension=2}{i4,v1,o1}
\fmf{vanilla}{i4,v1}
\fmffreeze
\fmf{phantom,tension=2}{i6,v2,v3,o6}
\fmf{phantom}{v1,v2}
\fmf{phantom,tension=2}{i2,v4,v5,o2}
\fmf{phantom}{v1,v4}
\fmffreeze
\fmf{vanilla,tension=2}{o7,v3,o5}
\fmf{vanilla,tension=2}{v1,v3}
\fmf{vanilla,tension=2}{o1,v5,o3}
\fmf{vanilla,tension=2}{v1,v5}
\end{fmfgraph}} \nonumber
\end{equation}
\end{fmffile}
Clearly
\begin{equation}
\label{F4_SYM_decompose}
F_4^{SYM}(\vk_1, \vk_2, \vk_3, \vk_4) = F_4^{1, SYM}(\vk_1, \vk_2, \vk_3, \vk_4) + F_4^{2, SYM}(\vk_1, \vk_2, \vk_3, \vk_4) \, .
\end{equation}

%%%%%%%%%%%%%%%%%%%%%%%
\subsection{Structure of higher order terms}
%%%%%%%%%%%%%%%%%%%%%%%

We claim that arrangements of the terms on the symmetrization process can be made at all orders in such a way that the symmetry properties of the subkernels is improved. 

For a given $n$ the recurrence relations produce all the possible repartitions  $(a_1 \ldots a_k)^{\pi} (a_{k+1} \ldots a_n)^{\pi}$ where the $a$'s represent the indexes of the momenta. If $n$ is odd,
the structure of the recurrence relations imply the existence of terms
$(b_1 \ldots b_{n-k})^{\pi}(b_{n-k+1} \ldots b_n)^\pi$. Since all
permutations must be present on the symmetrized kernels, it is always
possible to make the pairwise association of terms for which $a_1
\leftrightarrow b_{n-k+1}, \ldots, a_k \leftrightarrow b_n, a_{k+1}
\leftrightarrow b_1, \ldots a_n \leftrightarrow b_{n-k}$. Proceeding in this way it is possible to built, order by order, subkernels that are symmetric under the exchange of the first $k$ and last $n-k$ arguments, that will be, in our notation, the tilded subkernels. The full symmetrized kernel at a given order will be the sum of all tilded subkernels divided by the number of such objects. 

The exception happens if $n$ is even. In this case the repartition $(a_1 \ldots a_{n/2})^\pi (a_{(n/2+1)} \ldots a_m)^\pi$ only appears once on the list of all repartitions, and therefore there are no alternative pairings. These terms correspond to a specular diagrams, as observed in the case of $\tilde{F}_4^{2}$.

%%%%%%%%%%%%%%%%%%%%%%%%%%%%%%%%%%
\section{Recurrence relations for tilded kernels}
%%%%%%%%%%%%%%%%%%%%%%%%%%%%%%%%%%

Collecting our findings, we can rewrite the recurrence relation for $\tilde{F}_n$ given in Eq. \eqref{F_recurrence} as:

\begin{eqnarray}
\tilde{F}_n (\vk_1, \ldots, \vk_n) & =: & \sum_{m=1}^{\lfloor n/2
                                         \rfloor}
                                         \tilde{F}_n^{m}(\vk_1,
                                         \ldots, \vk_n) \nonumber\\ &
                                                                      =
                                       & \sum_{m=1}^{\lfloor n/2 \rfloor} \frac{1}{(2n+3)(n-1)} \Bigg\{ G_m^{SYM}(\vk_1, \ldots \vk_m) \Bigg[ (2n+1) \alpha(\bar{\vk}_1, \bar{\vk}_2) F_{n-m}^{SYM} (\vk_{m+1}, \ldots, \vk_n) \nonumber\\ & & + 2 \beta(\bar{\vk}_1, \bar{\vk}_2) G_{n-m}^{SYM}(\vk_{m+1}, \ldots, \vk_n) \Bigg] \nonumber\\ & & + \left[1 - \delta_{m, n/2}^K \right] G_{n-m}^{SYM}(\vk_{m+1}, \ldots, \vk_n) \Bigg[(2n+1) \alpha(\bar{\vk}_2, \bar{\vk}_1) F_m^{SYM}(\vk_1, \ldots, \vk_m) \nonumber\\ & & + 2 \beta(\bar{\vk}_2, \bar{\vk}_1) G_m^{SYM}(\vk_1, \ldots, \vk_m) \Bigg] \Bigg\} \, .
\end{eqnarray}
As defined before, $\bar{\vk}_1=\vk_1 + \ldots + \vk_m$ and
$\bar{\vk}_2 = \vk_{m+1} + \ldots + \vk_n$. Here $\delta^K_{m, n/2}$
is the Kronecker delta and corrects for the counting when $n$
is even.

In terms of the subkernels $\tilde{F}_n^m$, $F_n^{SYM}$ is written as:
\begin{equation}
\label{F_n_tilde_SYM}
F_n^{SYM}(\vk_1, \ldots, \vk_n) = \sum_{m=1}^{\lfloor n/2 \rfloor} \left( \begin{array}{c} n \\ m \end{array} \right)^{-1}  \tilde{F}_n^m (\vk_1, \ldots, \vk_n)^\pi \, ,
\end{equation}
where $\tilde{F}_n^m (\vk_1, \ldots, \vk_n)^\pi$ stands for the sum of
the  $\left( \begin{array}{c} n \\ m \end{array} \right)$ remaining
permutations for each class of subkernels. 

For $\tilde{G}_n$ the recurrence relation reads:
\begin{eqnarray}
\tilde{G}_n (\vk_1, \ldots, \vk_n) & =: & \sum_{m=1}^{\lfloor n/2
                                         \rfloor}
                                         \tilde{G}_n^{m}(\vk_1,
                                         \ldots, \vk_n) \nonumber\\& = & \sum_{m=1}^{\lfloor n/2 \rfloor} \frac{1}{(2n+3)(n-1)} \Bigg\{ G_m^{SYM}(\vk_1, \ldots \vk_m) \Bigg[ 3 \alpha(\bar{\vk}_1, \bar{\vk}_2) F_{n-m}^{SYM} (\vk_{m+1}, \ldots, \vk_n) \nonumber\\ & & + 2 n \beta(\bar{\vk}_1, \bar{\vk}_2) G_{n-m}^{SYM}(\vk_{m+1}, \ldots, \vk_n) \Bigg] \nonumber\\ & & + \left[1 - \delta_{m, n/2}^K \right] G_{n-m}^{SYM}(\vk_{m+1}, \ldots, \vk_n) \Bigg[3 \alpha(\bar{\vk}_2, \bar{\vk}_1) F_m^{SYM}(\vk_1, \ldots, \vk_m) \nonumber\\ & & + 2 n \beta(\bar{\vk}_2, \bar{\vk}_1) G_m^{SYM}(\vk_1, \ldots, \vk_m) \Bigg] \Bigg\} \, .
\end{eqnarray}
Also for $G_n^{SYM}$,
\begin{equation}
\label{G_n_tilde_SYM}
G_n^{SYM}(\vk_1, \ldots, \vk_n) = \sum_{m=1}^{\lfloor n/2 \rfloor}  \left( \begin{array}{c} n \\ m \end{array} \right)^{-1}  \tilde{G}_n^m (\vk_1, \ldots, \vk_n)^\pi \, .
\end{equation}

We observe that 
\begin{equation}
\label{growth_order}
\sum_{m=1}^{\lfloor n/2 \rfloor} \left( \begin{array}{c} n \\
                                          m \end{array} \right) =
                                      2^{n-1} -1 - \frac{1}{2}
                                      \left( \begin{array}{c} n\\
                                               n/2 \end{array} \right)
                                           \left[ n \, \mathrm{mod} \,
                                             2 - 1 \right] \, ,
\end{equation}
and therefore the number of terms in Eqs. \eqref{F_n_tilde_SYM}, \eqref{G_n_tilde_SYM} grows exponentially whereas in Eqs. \eqref{f_SYM}, \eqref{g_SYM} the number of terms has factorial growth. The computational time for high order kernels would be greatly reduced using the modified recurrence relations proposed here. 

It was  also discussed in \cite{bertolini2016non} a more efficient way
to compute the PT kernels. The symmetrization there
would correspond to the parallel
linking of terms in our discussion, i.e., leads to $F^s_n, G^s_n$. The
symmetrization cost for these kernels also grows as in
Eq. \eqref{growth_order}, but there is no permutation symmetry among $m$ firsts
and $n-m$ last arguments in each subkernel in this case. We claim 
that the tilded subkernels may therefore be better
suited when performing angular integrals, because the structure of
their denominators separate some of the dependences, what may lead to
better numerical performance. 

%%%%%%%%%%%%%%%
\section{Conclusion}
%%%%%%%%%%%%%%%

The evolution of initial density contrast and velocity fields for the dark matter fluid as described by continuity, Euler and Poisson equations should reproduce the development of gravitational instabilities that has led to structure formation in the Universe. Perturbation Theory provides kernels that relate all orders on the perturbative development of density and velocity fields to the linearly evolved density contrast. These kernels are functions of momenta, and only their component symmetric under the permutation of arguments can contribute to physical quantities, what motivates discussions on symmetrization procedures.

It follows from the basic fluid equations that continuity equation couples density and velocity, and Euler's equation couples velocity to velocity. When translated to Fourier space, the coupling structure is encoded in the functions $\alpha (\vk_1, \vk_2)$, that is associated continuity equation, and $\beta(\vk_1, \vk_2)$ that is associated to the couplings in Euler's equation. The source of asymmetry of the PT kernels is the asymmetry of $\alpha$: $\alpha (\vk_1, \vk_2) \neq \alpha (\vk_2, \vk_1)$, what impacts the recurrence relations that generate the kernels.

One simple way of symmetrizing the kernels is by summing over all possible permutation of momenta, and dividing the sum by the number of permutations. An alternative symmetrization approach can be defined by symmetrizing the vertex couplings in order to compensate for the non-symmetric nature of $\alpha$, and then sum over permutations of the remaining asymmetric factors. The second procedure leads to subkernels with better symmetry properties, and we have shown here how to recover such subkernels from the standard symmetrization scheme. The key point is the definition of the tilded kernels $\tilde{F}_n$, $\tilde{G}_n$, whose symmetry properties also imply better structure of momenta dependence in the kernels denominators. %The better analytical behavior allows to simpler calculation of certain angular integrals, and an application is the investigation of multipole decomposition of vacuum expectation values \cite{multipole}.

We provided recurrence relations in terms of which the tilded kernels
of a given order can be constructed from the full symmetrized version
of the kernels of all smaller orders. We should remark that the
symmetrization of the tilded kernels require a sum of terms that grows
at most exponentially, while the usual kernels have a symmetrization
cost with factorial growth. We emphasize, finally, that the tilded kernels emerge naturally from
Scoccimarro's method with symmetric vertex couplings.

\subsection*{Acknowledgments}
The author thanks Francis Bernardeau for useful discussions. This work
was supported by Grant No. ANR-12-BS05-0002 and the Labex ILP (Grant
No. ANR-10-LABX-63), part of the Idex SUPER of the Programme
Investissements d'Avenir under Grant No. ANR-11-IDEX-0004-02.

%%%%%%%%%%%%%%%%%%

\bibliographystyle{unsrt}
\bibliography{ref}

\end{document}